\documentclass[journal=jacsat,manuscript=article]{achemso}

\usepackage{chemformula} 
\usepackage[T1]{fontenc} 
\usepackage[utf8]{inputenc}
\usepackage{textcomp}

\author{Zahra Ahmadian Dehaghani}
\email{zahmadia@sissa.it}
\affiliation[]
{Scuola Internazionale Superiore di Studi Avanzati (sissa), Trieste, Italy}

\title[An \textsf{achemso} demo]
{Threading in star catenanes: The role of ring rigidity, topology and environmental crowding}

\abbreviations{IR,NMR,UV}

\begin{document}






\begin{abstract}
 This study investigates the probability of threading in star catenanes under good solvent conditions using molecular dynamics simulations, emphasizing the influence of ring rigidity. Threading in these systems arises from the interplay between the intrinsic topology of and within the star-shaped structure and the bending rigidity of individual rings. It is demonstrated that reduced ring flexibility enhances threading, and the presence of mechanical bonds is critical for threading formation. Notably, the bending rigidity of the rings alters their shapes, resulting in a non-monotonic threading probability with a peak at intermediate rigidity. Furthermore, increasing ring length is found to significantly boost threading probability. These findings elucidate the intricate relationships among topology and rigidity in governing threading, with implications for the design of advanced molecular systems and materials. This work provides a comprehensive framework for understanding threading in good solvent conditions, where such behavior is typically improbable for ring polymers, and opens avenues for the development of molecular machines and other complex architectures.

  
\end{abstract}

\section{Introduction}

Chains, the most basic topology in polymer science, consist of monomers connected together in a linear fashion. They have been extensively studied over decades, creating the foundation of polymer research. Rings, which are formed by closing the ends of a chain, and stars, where multiple chains radiate from a central core, represent other fundamental polymer topologies that have received considerable attention subsequent to the chains \cite{Hoagland-1997,rubinstein2003polymer,McLeish-1999}. Another interesting form of polymer topology emerges when rings are interlocked via mechanical bonds to build macromolecules with more complex topology known as catenanes \cite{GilRamirez-2015,TUBIANA-2024}. Studies on these topologies reveal that polymer architecture significantly influences the overall distribution of monomers and the size of the polymers, leading to a significant impact on physical properties such as viscosity, diffusion coefficient, and even their theta temperature \cite{SIKORSKI_1994,grosberg-1988,Wu_2017,Halverson-2012,Vargas-2018,Ahmadian-2020,Farimani-2024,Chen-2024,Luca_2022}.
 
Looking deeper into polymer dynamics reveals that entanglements, such as knots and threading, contribute complexity to the narrative. Knots form when chains become entangled with themselves, leading to a change in their size \cite{Dai-2016}. Studies have investigated the dynamics of knotting and unknotting in polymers, shedding light on the kinetics of knot formation and the factors that influence knot stability \cite{Ruskova-2023,Tubiana-2013,suma-2015}. Rigidity adds an extra layer of complexity and results in a non-monotonic behavior for probability of knotting in  ring polymers that arises from the competition between entropy cost of knotting formation and bending energy \cite{Coronel-2017-Softmatter}.

While knots can form in dilute solution in a single polymer, there exists another type of entanglement called threading that requires a population of polymers. Ring threading typically manifests in dense solutions of polymers and melts, wherein a polymer, such as chain or ring, penetrates a ring polymer \cite{WANG-2024,rosa-2020}. This process significantly prolongs relaxation time and slow down the dynamics of the system, often leading to the transition of a ring melt into a glassy state \cite{Tu-2023}.

Inter-ring threading happens when both threading and threaded polymers are rings. The study of inter-ring threading was initially explored through the diffusion of rings within a background gel, which provides an environment conducive to identifying and quantifying these interactions \cite{Micheletto-2014-Macroletter}. The investigation revealed that as the length of the rings increases, both the number and resistance time of the threading events rise, ultimately resulting in a percolation networks of threaded rings.

Presence of threading was later characterized by randomly pinning a fraction of rings in a concentrated solution, resulting in an induced arrested state. Threading events in such a system yield a novel glass transition which happens above the classic glass transition temperature\cite{Micheletto-2016-PNAs}. 

Further study reveals that ring polymers in melt, once threaded, are unable to diffuse until an unthreading event takes place \cite{Lee-Rapid}. Consequently, this phenomenon leads to a deceleration in the diffusion process of ring polymers and significant slow downs in the overall dynamics, with the decoupling between internal structure relaxation and diffusion being crucial for understanding this effect. 

In addition, the concept of ring minimal surface was suggested to quantify threading events in melts of ring polymers in which rings are well described as double-folded rings, showing less threaded conformation compared to their counterparts at equilibrium \cite{Smrek-2019-Macroletter,Chubak-2020,Ubertini-2022}.  

Furthermore, degree of threading, defined as the number of rings penetrated into a single ring, was explored in a melt of rings with various bending rigidities \cite{Guo-2020-polymers}. Similar to knot formation, bending rigidity results in a non-monotonic behavior for probability of threading for a given ring topology as the rigidity of rings varies, with a peak in intermediate regime where there is a balance between coil expansion and stiffness.

Recent observations have unveiled threading phenomena within a solitary linear catenane under poor solvent conditions \cite{Ahmadian-2023}. A linear catenane comprises interlocked ring polymers arranged in a linear configuration. The conditions of a poor solvent facilitate their close proximity and formation of threading since a population of rings are available in the compact state.

Although threading significantly influences the dynamics of the system, it is often overlooked in polymer theories, leading to their failure when threading is present within the system. In order to formulate threading in existence theories, it is essential to have a thorough understanding of its nature. Thus far, threading has been investigated in highly dense system of polymers where a given ring within the system is surrounded by a compact population of other rings. This raises the question of whether threading can also occur in less dense polymer systems in a good solvent.

This study investigates a star catenane system composed of multiple arms, each formed by linear catenanes interlocked with a central ring, using molecular dynamics simulations. This catenated structure ensures the proximity of rings within the system, while mechanical bonds between interlocked rings provide the freedom for them to move relative to one another. It is ensured that all rings within the system were of identical size and composed of the same number of monomers. This approach was adopted to eliminate potential asymmetry that might lead to unequal probabilities of rings serving as a host or guest \cite{Stano-2023-Softmatter}.

To understand the role of structural properties in threading, the stiffness of the rings is varied, altering their shape, and its effect on the probability of threading is analyzed. The star catenane design enables examination of both local and global topologies on threading behavior. Locally, threading within individual arms composed of linear catenanes is explored, while globally, the influence of the overall star structure on threading interactions is assessed. These insights shed light on the formation of threading events between linear catenanes maintained in close proximity within a good solvent, providing a foundation for understanding the interplay of rigidity, topology, and threading in mechanically interlocked systems.

\section{Model and Methods}
\subsection{Model Definition and Simulation Setup}

\begin{figure}
    \centering
    \includegraphics[width=0.8\textwidth]{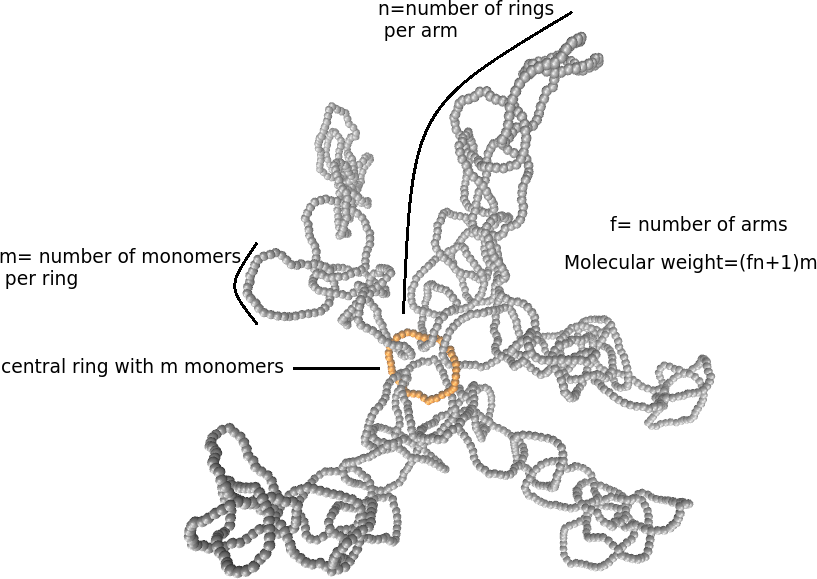}
    \caption{Equilibrium configuration of a star catenane with $m=40$, $f=5$, $n=10$ and $k_b=10.0$, displayed with five visible arms. Farther monomers are represented with brighter colors to enhance depth perception. The central ring is highlighted in orange to distinguish it from the rings within arms.  }
    \label{fig:model}
\end{figure}

The system considered in this work is a star catenane characterized using three parameters of $f$ arms, $n$ rings per arm and $m$ monomers per ring. All rings in each arm are interlocked via mechanical bonds and all arms are interlocked by mechanical bond to a central ring with the same number of monomers as other rings within the arms (Fig \ref{fig:model}). The star catenane overall has $N_r=f \times n+1$ rings and $N_m=N_{ring} \times m$ monomers. 

The reference system contains $f=5$ arms, $n=10$ rings in each arm, $m=40$ monomers per ring and total number of monomers $N_m=2040$. To gain better insight into the impact of these parameters on the probability of threading, various star catenanes are considered, with only one parameter varied while the other two parameters are kept fixed at the values of the reference system.

With this in mind, number of arms adapts the values of $f=3, 5, 7$ and $9$, the number of the rings per arm $n=5, 10, 15, 20$ and $25$ and the number of monomers per ring $m=20, 40, 60$ and $100$. The total number of monomers varies from $N_m=1020$ for the smallest system with $m=20$ to $N_m=5100$ for the biggest one with $m=100$.

All monomers interact with each other via truncated and force-shifted Lennard-Jones potential in which standard Lennard-Jones function is subtracted by a linear function, ensuring that both potential and force are continuous at the cut off of $r_c=\sqrt[6]{2} \sigma$. $\sigma=1.0$ is the diameter of beads and it is chosen as the length unit. The function of $\phi(r)$ as the Lennard-Jones potential is considered as below:
\begin{equation}
\phi\left(r\right) =  4 \epsilon \left[ \left(\frac{\sigma}{r}\right)^{12} -
                    \left(\frac{\sigma}{r}\right)^6 \right]
\end{equation}
In which $\epsilon=1 K_B T$ is the depth of the potential well and is considered as the energy unit. $K_B$ represents Boltzmann constant and $T$ is the temperature of the system. The parameter $r$ shows the distance between two interacting monomers. Truncated and force-shifted Lennard-Jones potential is formulated as:
\begin{equation}
U_{LJ}\left(r\right)=
\begin{cases}
\phi\left(r\right)  - \phi\left(r_c\right) - \left(r - r_c\right) \left.\frac{d\phi}{d r} \right|_{r=r_c}       & r <= r_c \\
0 & r > r_c
\end{cases}
\end{equation}
Where $\frac{d\phi}{d r}$ is the derivative of $\phi(r)$ with respect to $r$. 

Each two bonded consecutive monomers maintain connectivity via finite extensible nonlinear elastic (FENE) potential defined as:

\begin{equation}
U=
\begin{cases}
    -0.5 K R_0^2  \ln \left[ 1 - \left(\frac{r}{R_0}\right)^2\right] + 4 \epsilon \left[ \left(\frac{\sigma}{r}\right)^{12} - \left(\frac{\sigma}{r}\right)^6 \right] + \epsilon & r <= R_0 \\
    \infty & r > R_0
\end{cases}
\end{equation}
In which $R_0=1.5 \sigma$ and $K=30 \epsilon /\sigma^2$. The stiffness of the rings within the structure is varied by a cosine potential defined as below:
\begin{equation}
    U_b = k_b [1 - \cos(\theta - \theta_0)]
\end{equation}
Where $\theta$ is the angle between two neighboring bond vectors in a ring and $\theta_0=180^\circ$ is chosen as equilibrium value for the defined angle. $K_b$ represents the stiffness of the ring and is proportional to the persistence length of a linear polymer analogue. The stiffness constant, $K_b$, varies in the range $K_b/m = 0.0 \, \epsilon/\sigma^2$ to $K_b/m = 1.0 \, \epsilon/\sigma^2$ for a given $m$.
To facilitate comparison, a system with one single isolated ring with $m=40$ is also considered.

Molecular dynamic simulation is performed using LAMMPS package \cite{LAMMPS} at the temperature of $T=1 K_B T /\epsilon$ by applying Langevin thermostat with a standard friction coefficient \cite{kremer_1990}. The equation of motions were integrated by the time step of $\Delta t=0.005 \tau_{LJ}$, where $\tau_{LJ}=\sigma \sqrt{m/\epsilon}$ with $m=1$ as mass of a single bead chosen as mass unit. For each set of parameters for star catenane, data is gathered from $5-10$ independent runs with the duration of $4 \times 10^{+6} - 6 \times 10^{+6} \tau_{LJ}$. The values of observables are averaged over all available independent runs and error bars are calculated using standard deviation of the averages taken from each run.

\subsection{Observables}
\subsubsection{Size and Shape}
Both size and shape of a ring polymer can be characterized by gyration tensor which is defined as:
\begin{equation}
G_{\mu \nu}=\frac{1}{2N^2} \Sigma_{i=1}^{i=m} \Sigma_{j=1}^{j=m}(r^{\mu}_i-r^{\mu}_j)(r^{\nu}_i-r^{\nu}_j)
\end{equation}
Where $m$ is the total number of monomer in the polymer and $\mu,\nu=x, y$ and $z$. The eigen values of the gyration tensor measure the spatial extension of polymer along principle axis and are denoted by $\lambda_k$ with $k=1,2$ and $3$ in a descending order $\lambda_1 \geq \lambda_2 \geq \lambda_3$. The size of the polymer is measured based of radius of gyration, $R_g$, defined as $R_g^2=\lambda_1^2+\lambda_2^2+\lambda_3^2$ \cite{eigen}.

To characterize the shape of a polymer, asphericity, $\Delta$ and nature of asphericity are measured, $\Sigma$ \cite{Alim-2007}. Deviation from a fully symmetric sphere is given by $\Delta$ as below:
\begin{equation}
\Delta=\frac{3}{2} \frac{Tr \hat{G}^2}{(TrG)^2}
\end{equation}
Where $\hat{G}_{ij}=G_{ij}-\delta_{ij} \frac{TrG}{3}$ and $\Delta$ is bounded between $0 \leq \Delta \leq 1$. While $\Delta=0$ is obtained for a fully symmetric polymer, $\Delta=1$ is a result of fully extended polymer in form of a rod shape. 

Prolatness and oblateness of the polymer is specified by $\Sigma$ where it takes the values $-1 \leq \Sigma \leq 1$. For $\Sigma=-1$, the polymer is oblate, forming a structure similar to a disc and $\Sigma=1$ corresponds to a prolate shape.

\subsubsection{Threading Identification}

\begin{figure}
    \centering
    \includegraphics[width=1\textwidth]{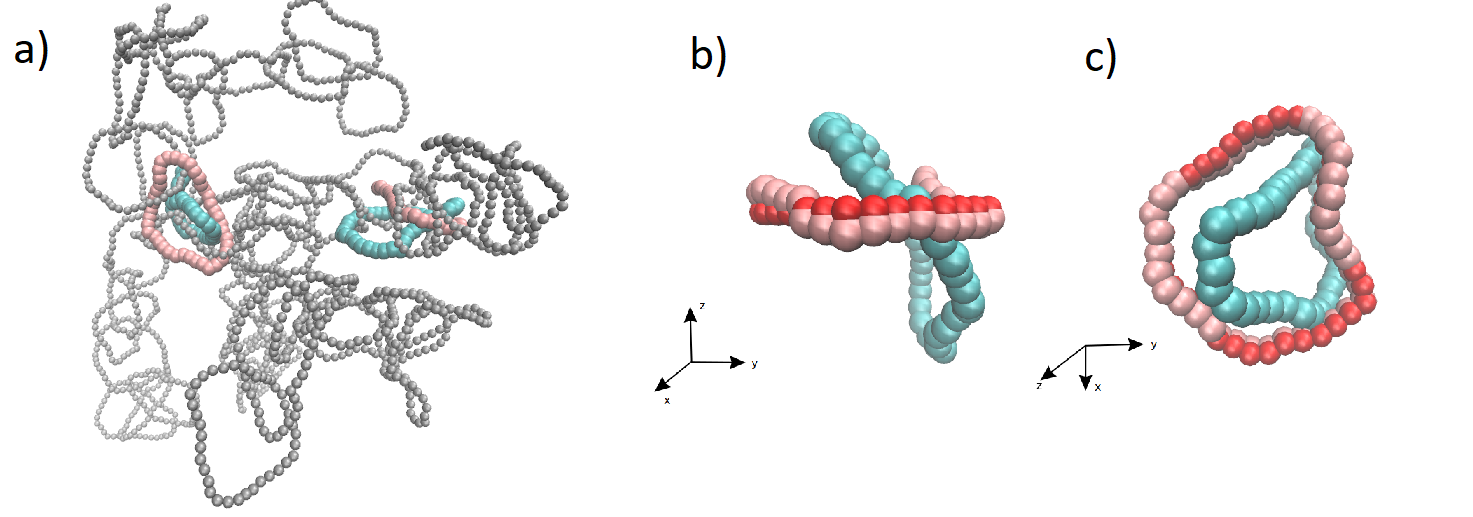}
    \caption{ 
(a) A threaded configuration of a star catenane (reference system) with two threading events, where the guest is shown in blue and the host in pink. (b) Illustration of threading events, showing the host, guest, and the projected host (red) for clarity. (c) The same threading events as in (b), viewed from another angle, demonstrating that the projected host provides a good approximation of the real host. All cases correspond to \( k_b = 10.0  \, \epsilon/\sigma^2\). 
}
    \label{fig:threading}
\end{figure}

An illustration of a star catenane threaded configuration is shown in Fig \ref{fig:threading}. In a single threading event shown in \ref{fig:threading} (b) and (c), The threaded ring (pink ring) is labeled as a "host" and the threading one (blue ring) as a "guest". Threading of ring I (as host) by ring J (as guest) can be identified as follow: First, ring I is projected in the plane made by the two of its eigen vectors which are equivalent to the biggest eigen values of its gyration matrix. In the next step, the projected shape is approximated with an ellipse centered in its center of mass and with the vertex and the co-vertex located in the position of monomers with maximum distance from the center of mass of the projected shape along eigen vectors.

Threading is identified if the number of intersections of ring J  with the ellipse surface is at least $2$. However, with this criterion, ring J is identified as guest in unthreaded cases that it only touches the surface of the ellipse. To discard this unwanted cases, a second criterion is considered which is applied on the depth of threading. This quantity is defined as the minimum of maximum distance of ring J's monomers from ellipse plane. Considering these two criteria, threaded configurations are identified by accuracy of $98.99 \%$ for $K_b=3.0 \, \epsilon/\sigma^2$ and $100.0 \%$ for $K_b=40.0 \, \epsilon/\sigma^2$.   

 

\section{Results and Conclusion}

\subsection{Probability of threading}

The probability of threading, $P_{th}$, is calculated by tallying the occurrences of threaded configurations for a star catenane. A threaded configuration in context of a star catenane is a trajectory frame wherein at least two rings are threaded in the whole configuration. A threaded configuration might contain more than one threading events. Threading events between two neighboring rings interlocked via mechanical bonds are excluded from consideration. It is important to note that threading events observed in this work are transient, and no long-lived occurrences were observed throughout the simulations for various parameters. Moreover, in all instances of threading, each host accommodated only a single guest, with no cases of threading involving two guests being observed.

All detected threaded events are further classified into following categories: (i) intra-arm threading, where host and guest belong to the same arm, (ii) inter-arm threading, featuring host and guest from two distinct arms and (iii) central threading, wherein either guest or host is the central ring, implying involvement of the central ring in a threading event.   

First, a single isolated star catenane with $f=5$ arms, $n=10$ rings per arm with various $m=20, 40, 60, 100$ monomers per rings including the central ring is considered and the probability of threading is calculated while ring's bending rigidity per monomer, $K_b/m$, was varied from $0.0$ for fully flexible rings to $1.0$ where persistence length (proportional to $k_b$) is equal to the contour length of the ring.

\begin{figure}
    \centering
    \includegraphics[width=1\textwidth]{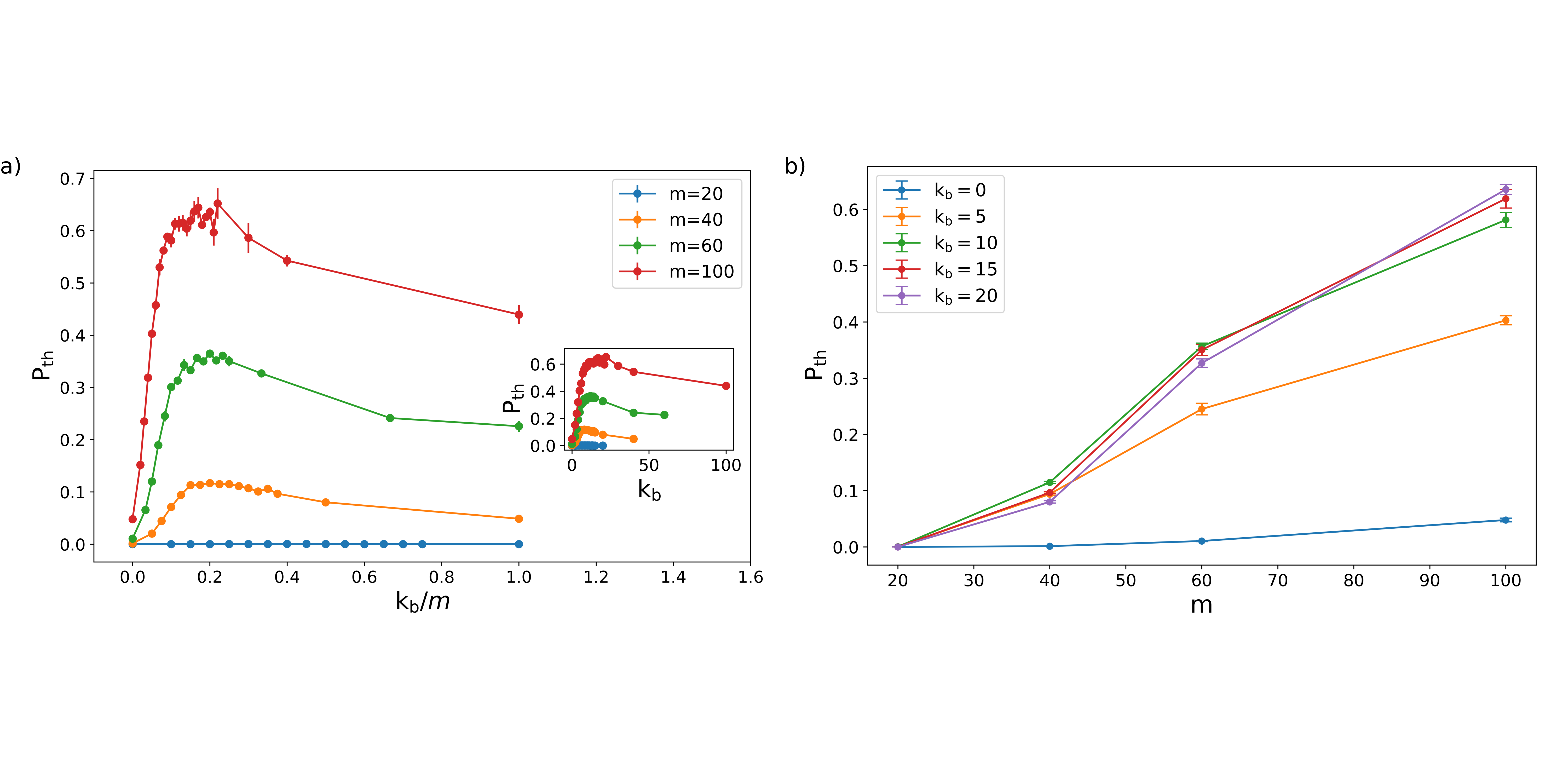}
    \caption{ 
Probability of threading for star catenanes with various \(m = 20, 40, 60, 100\), \(f = 5\), and \(n = 10\).  
(a) Non-monotonic behavior of probability of threading as a function of \(k_b/m\) for better comparison. The inset shows the probability as a function of \(k_b\).  
(b) Probability of threading versus \(m\) for selected \(k_b\) values across different flexibility ranges. Error bars are smaller than the dot size.  
}
    \label{fig:Prob_m}
\end{figure}

A non-monotonic behavior was observed for the probability of threading as shown in Fig \ref{fig:Prob_m} (a) with the change of rings' stiffness. Considering star catenane with $m=40$, it was observed that at first, $P_{th}$ grows significantly with $K_b$ and reaches a maximum value for intermediate stiffness and then, it declines gradually as the rings becomes more rigid. This trend is more evident for $m=60, 100$ with a peak in intermediates stiffness. For the case of $m=20$, probability of threading was found to be negligible. It suggests that threading requires sufficiently large rings to occur and is rarely observed for m=$20$, even with the change of $K_b$.

Threading is favorable when a guest can compress and penetrate into a host while the host can expand and increase the accessible space for the guest to pass through. As bending rigidity increases, the rings become more favorable for being a host but simultaneously less favorable for being a guest. This competition between the conformational preferences of the rings for being a host and a guest, governed by bending rigidity, results in a non-monotonic behavior for probability of threading.

The findings in Fig. \ref{fig:Prob_m} (a) establish the primary outcome of the study, affirming that increasing stiffness of rings yields a non-monotonic trend for probability of threading. The effect emerges as a consequence of bending rigidity, as evidenced by analogue observations in other polymer entanglements, such as threading in dense ring solution \cite{Guo-2020-polymers} or knot formation in a ring polymer\cite{Coronel-2017-Softmatter}.

 It is evident from Fig \ref{fig:Prob_m} (a) that for a given $K_b$, threading is more probable as the size of the ring increases. As shown for $m=100$, for $k_b \geq 5.0 \, \epsilon/\sigma^2$ over $40.0 \%$ of configurations are threaded even for the most rigid case studied in this work. 

This indicates that probability of threading can increase significantly with changes in the size of the rings. This trend can be attributed to the fact that larger rings provide more accessible space for the host, facilitating the penetration of the guest through it. However, when $m$ increases, the size of the guest also increases, which might intuitively suggest that threading becomes more challenging for larger and more rigid rings. Nevertheless, Fig \ref{fig:Prob_m} (b) demonstrates otherwise, showing that the probability of threading exhibits a monotonic behavior with increasing ring size $m$, even at high bending rigidity. 

Moreover, two additional observations are noteworthy: (i) $P_{th}$ is negligible in star catenane with fully flexible rings, with only $0.14 \%$ and $4.7 \%$ of configurations threaded respectively for $m=40$ and $m=100$, (ii) $P_{th}$ for the most rigid case, $K_b/m=1.0$, is $48$ times and $10$ times higher than a fully flexible one for $m=40$ and  for $m=100$, respectively. 

These observations establish a key result of the study. In good solvent, rigidity plays a crucial role in facilitating occurrence of threading. Flexible rings tend to adapt a crumpled configuration, limiting the space available for another ring to pass through. Introducing stiffness, however, causes the rings to open up, facilitating the threading process. While one may expect that high bending rigidity hinders threading, considering unfavorable compressing required for guest penetration through the host, the data demonstrate that stiff rings exhibits more favorability for threading compared to flexible ones.

\subsection{Metric properties of rings with different entanglements}

To investigate the underlying reason for observed behavior of $P_{th}$, metric properties of a single ring in four various states are examined as follows: (i) a typical concatenated ring within the star catenane structure interlocked via mechanical bonds to its neighboring rings labeled as "macrocycle", (ii) a single isolated ring without any mechanical bonds labeled as "isolated", (iii) host in a threaded state and (iiii) guest of the equivalent host. First, the impact of $K_b$ on their radius of gyration in the four mentioned states was considered. Despite sharing the same topology as a ring, they differ in their types of entanglements.

\begin{figure}
    \centering
    \includegraphics[width=1\textwidth]{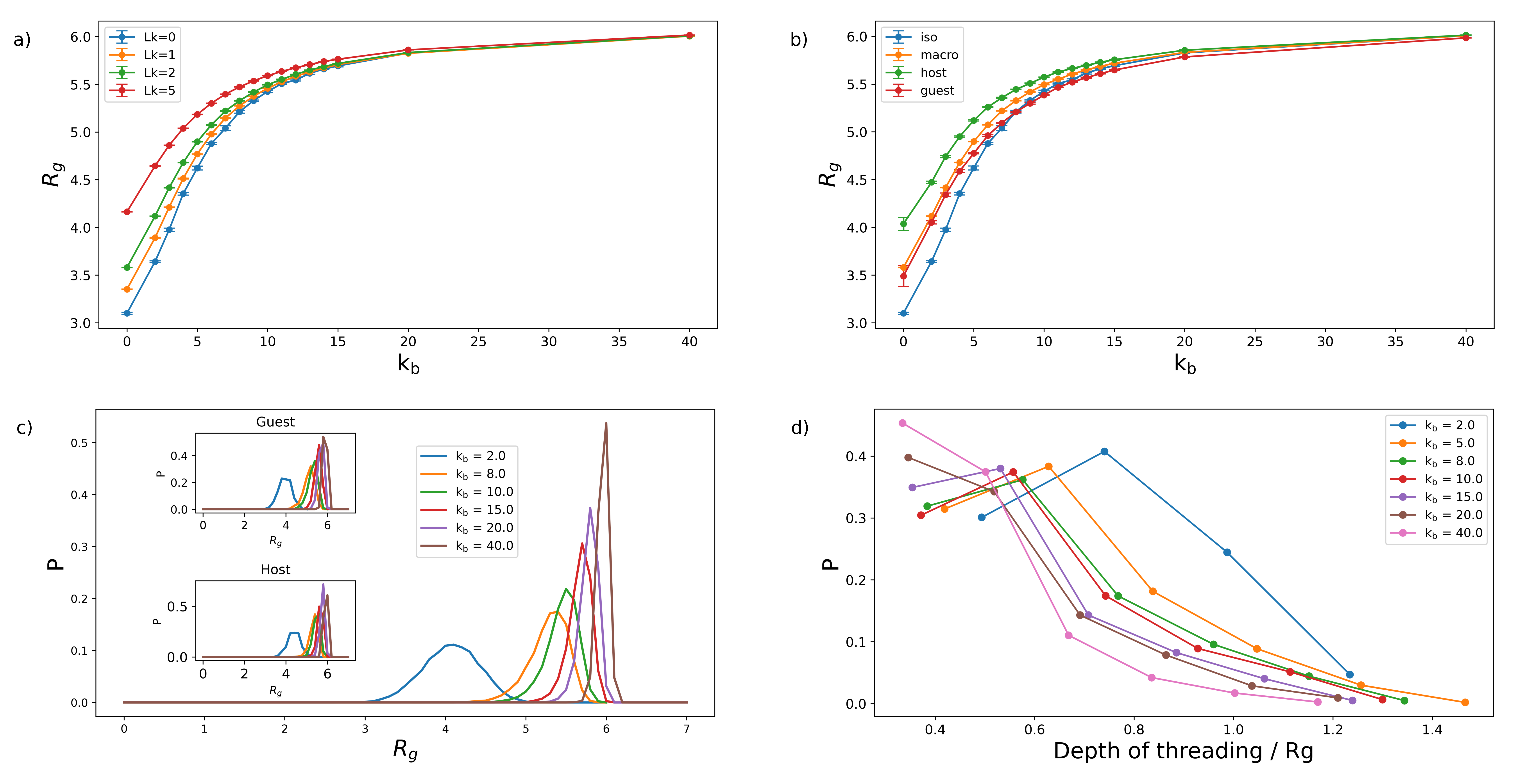}
    \caption{ 
(a) Comparison of \(R_g\) (radius of gyration) for a ring with various \(L_k\). The \(R_g\) does not saturate for rings with high bending rigidity.  
(b) Comparison of \(R_g\) for rings with different levels of entanglement, highlighting the impact of topological constraints on ring size.  
(c) Distribution of \(R_g\) for a macrocycle, host, and guest at selected \(k_b\) values.  
(d) Depth of threading as a function of stiffness, with the x-axis normalized to the \(R_g\) of the guest for better comparison across cases.  
}

    \label{fig:Rg}
\end{figure}

The results is shown in Fig. \ref{fig:Rg} for the reference system with $m=40$. For a given ring, parameter $L_k$ was defined as the number of interlocked rings associated with the ring in question. In this context, a single isolated ring has $L_k=0$, whereas all rings within a star catenane satisfy $L_k > 0$. Specifically, $L_k=1$ for the rings at the ends of each arm, $L_k=f$ for the central ring, and $L_k=2$ for the rings within each arm.

For an arbitrary bending rigidity within the considered range, it was observed that a macrocycle within the structure consistently adopts a larger value of $R_g$ compared to an isolated one. This is attributed to the presence of mechanical bonds within the structure. As shown in Fig \ref{fig:Rg} (a), $R_g$ adapts even higher values as the number of interlocked rings increases. This observation is consistent with previous findings in linear catenanes, which suggest that neighboring interlocked rings exert a pulling effect on the ring, resulting in an increase in its size \cite{Ahmadian-2020,Rauscher-2020}. However, at high bending rigidity, this effect disappears, and rings with various $L_k$ values tend to adopt similar sizes due to the lack of conformational freedom imposed by the bending rigidity.

For comparison, the "macrocycle" in this work refers to the rings within the structure with $L_k=2$, as the majority of macrocycles contain only two interlocked neighboring rings. As $K_b$ values ranges from $0$ to $40.0$, $R_g$ steeply rising in all cases at first and then, gradually is growing as $K_b$ reaches its maximum value of $40.0$. Even for $k_b=40.0 \, \epsilon/\sigma^2$, the radius of gyration is not saturated, indicating that the ring at the highest bending rigidity retain a degree of conformational flexibility and does not adapt a fully rigid circular structure.

For macrocycles in Fig \ref{fig:Rg} (c), it is shown that as bending rigidity increases, the $R_g$ distribution becomes more concentrated around its mean, resulting in a sharper peak. This behavior arises from the fact that more flexible rings can readily explore a wide range of conformations, from compact to extended, leading to a smaller mean for $R_g$ distribution. In contrast, increased bending rigidity restricts conformational flexibility due to the high energetic cost of bending, resulting in greater similarity among conformations and a sharper peak in the $R_g$ distribution. Furthermore, as the conformations become predominantly extended, the mean for $R_g$ also increases compared to that of more flexible rings.

To study the effect of entanglement on ring size, the radius of gyration ($R_g$) for the four mentioned states is shown in Fig. \ref{fig:Rg} (b). As observed, for a given stiffness, the radius of gyration for a guest is lower than that of a macrocycle, while for a host, it is significantly higher. This can be explained by the fact that during threading, the guest must compress to pass through the host, whereas the host must expand to provide accessible space for the guest to pass through. This observation demonstrates how threading, as an entanglement, alters the size of the rings. However, this effect diminishes as stiffness increases, as the influence of bending rigidity becomes dominant over the entanglement effect.

An intriguing observation arises from comparing the $R_g$ of isolated and guest rings for various $K_b$ values. For smaller values of $K_b$, the $R_g$ of the guest surpasses that of the isolated ring. However, around $K_b=8.0 \, \epsilon/\sigma^2$, this trend reverses, and the $R_g$ of the guest becomes slightly smaller than that of the isolated ring. It is noteworthy that two factors influence the radius of gyration of the guest compared to the isolated ring: (i) the presence of mechanical bonds, which tends to increase $R_g $, and (ii) the contraction required for the guest to penetrate into the host, resulting in a decrease in $R_g$ compared to the isolated ring. The interplay between these two factors leads to the reversal of the trend in the intermediate regime for the measured size of the guest and isolated ring.

This reversal of the trend in the guest ring's
size, where it becomes slightly smaller than the isolated ring, suggests a transition point in the balance between the mechanical bond-induced expansion and the contraction necessary for guest penetration. This reversal is likely a result of a critical threshold in bending rigidity, where the entropic effects of flexibility become dominated by the enthalpic penalty associated with bending energy.

The depth of threading in relation to bending rigidity was explored by examining the threaded states. As previously defined, the depth is the minimum of the maximum distance of the guest's monomers from the host's projected ellipse plane. The results for various bending rigidities are shown in Fig. \ref{fig:Rg} (d), with the depth normalized by the radius of gyration of the guest for clarity.

First, it is important to note that due to the restrictions applied in the algorithm for identifying threading, the distribution starts at a non-zero value, as zero depth is not recognized as a threaded event. Additionally, the case of fully flexible rings is excluded from the analysis due to insufficient threading observed in this case, which resulted in unreliable statistics.

For $K_b=20.0 \, \epsilon/\sigma^2$ and $K_b=40.0 \, \epsilon/\sigma^2$, most of the threading events are shallow due to the expanded configuration of the guest induced by high bending rigidity, which hinders penetration. As a result, the probability of deep threading decreases, making shallow threading more favorable. As the bending rigidity decreases and the rings become more flexible, a peak appears in the depth distribution at intermediate threading depths. This suggests that more flexible rings can penetrate deeper into the host, owing to the increased conformational freedom afforded by lower bending rigidity. However, it is noteworthy that this peak shifts to higher depths as flexibility increases. One might expect that for even more flexible systems, such as for $K_b=2.0 \, \epsilon/\sigma^2$, deep threading would be unfavorable due to the crumpled shape of the host, which reduces the available space for guest penetration. However, the results indicate otherwise, showing that deep threading becomes favorable for flexible rings. 

This counterintuitive result suggests that the flexibility of the guest ring allows it to more easily navigate the crumpled structure of the host, facilitating deeper penetration. The findings demonstrate a complex relationship between bending rigidity, flexibility, and threading depth. The guest ring's ability to penetrate the host depends not only on the available space but also on its own flexibility, emphasizing the critical role of conformational freedom in enabling deeper threading, with the flexibility of the guest being a key factor in facilitating this process.

\begin{figure}
    \centering
    \includegraphics[width=1\textwidth]{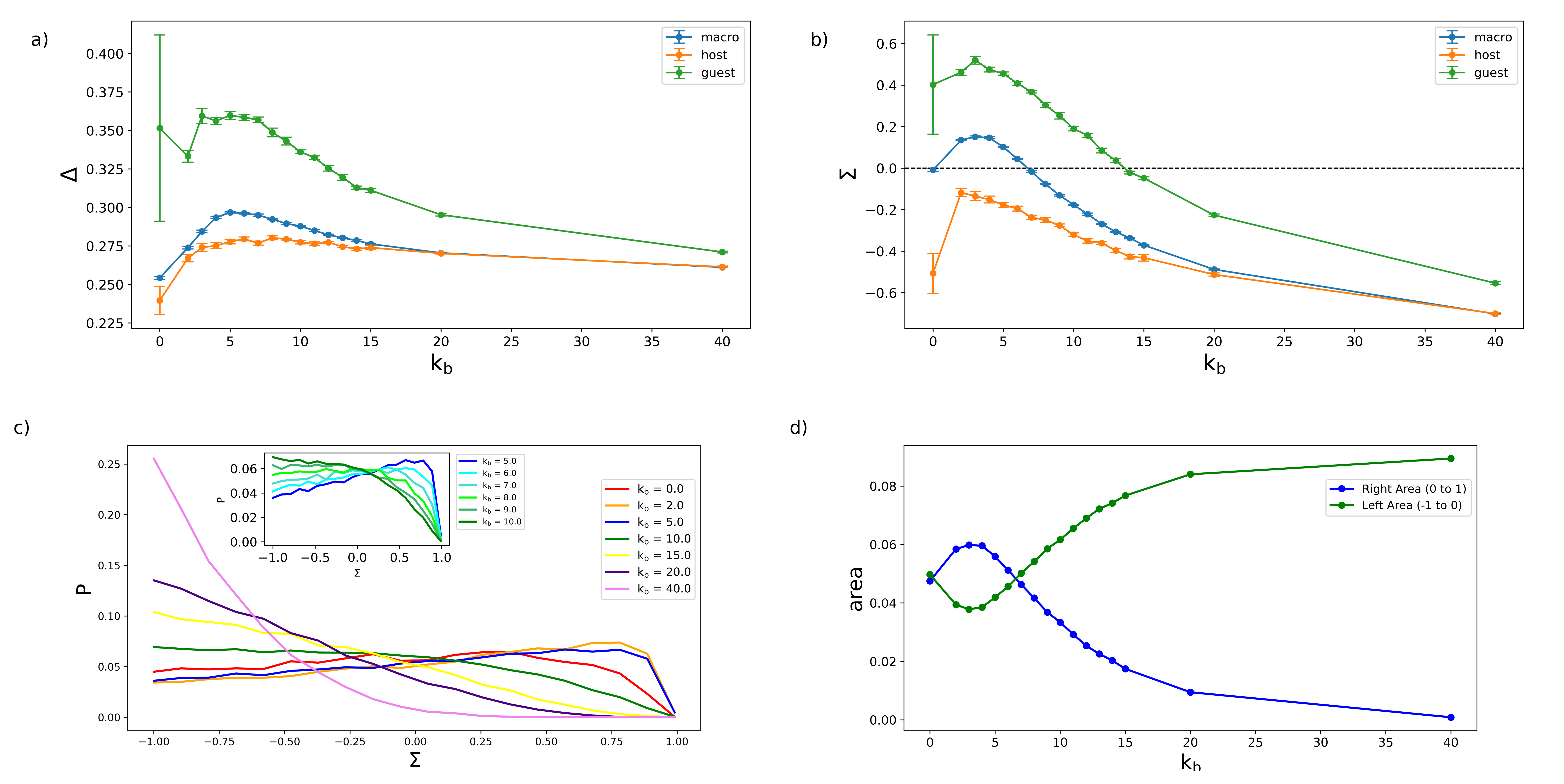}
    \caption{ 
(a) Comparison of \(\Delta\) for macrocycle, guest, and host at various stiffness values.  
(b) Non-monotonic behavior of \(\Sigma\) for macrocycle, guest, and host across various stiffness values. The dashed line at \(\Sigma = 0.0\) represents the point where positive and negative \(\Sigma\) values cancel each other out.  
(c) Distribution of \(\Sigma\) for the macrocycle at selected \(k_b\) values. The inset provides a clearer view of the intermediate regime.  
(d) Area under the distributions in (c) for positive values of \(\Sigma\) (right area) and negative values of \(\Sigma\) (left area). 
}

    \label{fig:Sigma}
\end{figure}

The asphericity (\(\Delta\)) and the nature of asphericity (\(\Sigma\)) provide critical insights into the shapes of the host, guest and macrocycle. These parameters, denoted as \(\Delta_i\) and \(\Sigma_i\) for \(i = \text{host, guest, macro}\), serve as quantitative measures of deviations from spherical symmetry and the anisotropic nature of the rings. Analysis of the data presented in Fig. \ref{fig:Sigma} reveals key observations that enhance our understanding of underlying mechanisms driving the non-monotonic behavior observed in the probability of threading. Ultimately, this analysis brings us closer to fully explaining the origin of this non-monotonic behavior.

As shown in Fig \ref{fig:Sigma} (a), the asphericity values satisfy the relation $\Delta_{guest} > \Delta_{macro} \geq \Delta_{host}$ for all values of $k_b$. Given that the asphericity of a perfectly symmetric sphere is $\Delta=0$ and a circular shape is $\Delta=1/4$, it was observed that fully flexible macrocycle and host rings tend to adopt a shape that is nearly symmetric and close to circular. However, as the bending rigidity increases, their shapes deviate from circular symmetry, becoming more distorted. In contrast, the guest ring exhibits consistently higher asphericity across all $k_b$ values, indicating a pronounced tendency to form more elongated and asymmetric structures regardless of bending rigidity.

For stiff rings in this study with $k_b=20.0 \, \epsilon/\sigma^2$ and $k_b=40.0 \, \epsilon/\sigma^2$, the macrocycle and host exhibit similar asphericity and nature of asphericity values, indicating that bending rigidity minimizes shape differences in these components. However, the guest retains a slightly more elongated shape relative to the host. This suggests that during threading, the effect of entanglement on the guest's shape persists despite stiffening of the rings.

While bending rigidity is a key factor in shaping the rings, the deviation in asphericity values for the host and guest compared to the macrocycle underscores the significant role of threading as a form of entanglement on the conformational structure of rings. This highlights the dual influence of bending rigidity and threading in determining the shapes of the threaded and threading rings.

By examining $\Sigma_i$, a richer insight was gained into the influence of threading on the shapes of the rings. For host, guest and macrocycle, $\Sigma_i$, initially exhibits a steep increase, reaching a maximum value, and subsequently decreases gradually. The negative values of $\Sigma_{host}$ across all bending rigidities suggest that the host consistently prefer an oblate shape ( Note that $\Sigma=-1$ corresponds to a flattened shape, such as a disk). On the contrary, upon increase of bending rigidity, guest initially form an elongated shape, corresponding to positive value for $\Sigma_{guest}$, before gradually flattened as $K_b$ approaches its maximum value (here, $\Sigma=+1 $ indicates an elongated object, such as rod).

Due to the compression of guest inside the host, it exhibits a greater degree of prolaticity, even for high bending rigidity, where both threaded and threading rings tend to form a disk-like shape under influence of bending rigidity. For $k_b <13.0 \, \epsilon/\sigma^2$, where the guest adopts an elongated shape and the host adopts an oblate shape, it can be concluded that macrocycles with positive $\Sigma$ values are more suitable as a guest, while those with negative $\Sigma$ values are better suited for the host.

To explore the above observation in more details, $\Sigma_{\text{macro}}$ offers intriguing insights as stiffness increases. As the macrocycles become less flexible, $\Sigma_{\text{macro}}$ transitions from positive to negative values in the intermediate regime with $\Sigma_{macro}=0.0$ at the transition point. Notably, $\Sigma_{\text{macro}}$ reaches a value of zero twice: once near $k_b = 0.0 \, \epsilon/\sigma^2$ for flexible rings, and again in the intermediate regime, for $5 \, \epsilon/\sigma^2 < k_b < 10b\, \epsilon/\sigma^2$ where the sign of $\Sigma_{macro}$ changes.

Here, $\Sigma$ values represents the mean of this quantity over a number of rings with a given bending rigidity, indicating that the rings adapt both positive and negative sigma values for $\Sigma_{macro}=0.0$. The cancellation of these values results in an overall zero, suggesting that macrocycles with a given bending rigidity exhibit diverse shapes. For flexible rings near $k_b=0.0 \, \epsilon/\sigma^2$, this behavior is expected, as they can easily adopt a variety of shapes due to their high flexibility. However, the occurrence of a zero value in the intermediate regime is more intriguing, as it suggests that even in this regime, where flexibility is reduced, the rings still exhibit a range of shapes that balance out to zero.

By looking at the distribution of $\Sigma_{macro}$, a clear illustration of this can be concluded. As shown in Fig for $m=40$, $\Sigma$ distribution is always asymmetric due to the topology of the ring, which prevents it from adopting a fully rod-like shape at high bending rigidity. For smaller $K_b$, the peak of distribution is close to $\Sigma_{macro}=1$, emphasizing an elongated shape is more probable, and macrocycles are more likely to serve as candidates for the guest rather than the host. However, for higher value of $K_b$, number of macrocyles with oblate shape are dramatically increasing, leading to uneven number of guest and host candidates in various $K_b$.

In the intermediate regime, there exists a value of \( k_b \) for which the number of rings with positive and negative values of \( \Sigma_{\text{macro}} \) are equal, leading to \( \Sigma_{\text{macro}}=0 \). To locate this point more accurately, the area under the probability distribution for both the positive and negative ranges of \( \Sigma_{\text{macro}} \) were calculated. It is observed that for intermediate values of bending rigidity, where $\Sigma_{macro}$ adapts a value close to zero, the macrocycle within the structure adopt various forms, with positive values of \( \Sigma_{\text{macro}} \) being more favorable for the guest and negative values more favorable for being a host.
This suggests that at intermediate values of \( k_b \), the number of available host and guest candidates for threading are comparable, leading to the maximum probability of threading or the highest number of threading events. This observation demonstrates how the formation of threading can be controlled by the stiffness of macrocycles. The competition between the favorability of macrocycles for being a host or guest, dictated by the ring's stiffness, gives rise to a non-monotonic behavior in the probability of threading as observed. 

It is important to note that this behavior does not apply to fully flexible rings, where $\Sigma_{macro}$ is close to zero, since the crumpled shape of flexible rings prevents threading from occurring. This highlights the essential role of bending rigidity in enabling threading formation.

\begin{figure}
    \centering
    \includegraphics[width=1\textwidth]{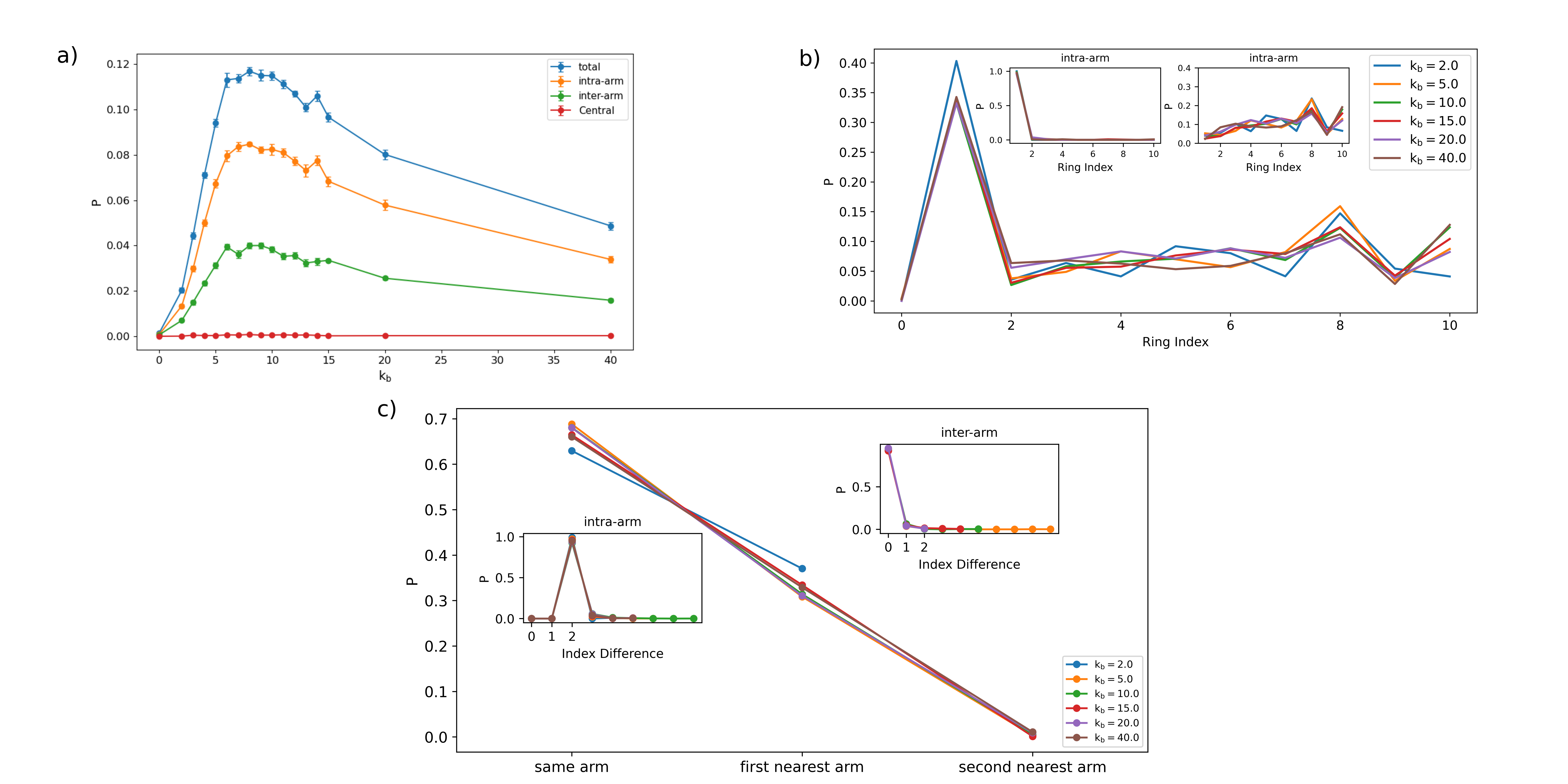}
    \caption{Probability of threading within and among the arms of a star catenane. (a) Threading probabilities are divided into three categories: intra-arm, inter-arm, and central threading, with central threading being negligible. (b) Probability of a ring acting as a host within one arm, showing that the first ring of each arm, interlocked with the central ring, has the highest probability. (c) Probabilities of inter- and intra-arm threading based on the relative positions of the arms accommodating the host and guest, categorized as same, first nearest, and second nearest arm. Subplots demonstrate the index differences between host and guest for intra- and inter-arm threading.}
    \label{fig:inter_intra}
\end{figure}

\subsection{Intra- and Inter-Arm threading in star catenanes}

Thus far, the analysis has treated the star catenane as an isolated catenated structure, focusing solely on the influence of rigidity and the mechanical bonds of its components on threading. In this section, a more detailed perspective, is adopted by viewing the star catenane as a system composed of multiple linear catenane radiating from a central ring. This framework allows us to explore the intra-arm and inter-arm threading, as defined previously, to gain deeper insights into threading behavior of this complex system.

Here, the reference system for star catenane, is considered again with $m=40$, $f=5$ and $n=10$. The total threading probability is decomposed into contributions from intra-arm and inter-arm and central threading. Fig \ref{fig:inter_intra} (a) shows that threading predominantly occurs within individual arms, although a significant fraction of inter-arm threading is also observed. Notably, the probability of threading through the central ring is negligible across all stiffness values.

Fig. \ref{fig:inter_intra} (b) provides a detailed analysis by illustrating the probability of each ring within the structure acting as a host. To quantify this, an index was assigned to each ring: the central ring is designated as $\text{macro}_0$, while within each arm, $\text{macro}_1$ represents the first ring interlocked with the central ring, progressing sequentially towards the end of the arm. The terminal ring at the end of each arm is indexed as $\text{macro}_n = \text{macro}_{10}$.

The distribution is calculated for all threadings, encompassing both inter-arm and intra-arm and central threading events. The subplots further separates contributions of intra-arm and inter-arm, highlighting that the ring interlocked with the central ring, $\text{macro}_1$, plays a significant role in inter-arm threading. This ring predominantly serves as the host for threading events between arms, emphasizing its crucial role in the contribution of other arms on threading.

The subplot for intra-arm interactions reveals a non-uniform distribution of the probability of a ring acting as a host, which increases as the ring indices increase from the center of the star toward the end of the arm. This trend can be attributed to the reduced steric crowding and increased flexibility of the rings farther from the central core, which enhances their ability to engage in threading events. Rings closer to the core experience limited movement due to the crowding of neighboring rings, whereas as the end of the arm is approached, this crowding diminishes, allowing the rings greater freedom of motion and flexibility, which facilitates their involvement in threading.

An unexpected behavior is observed at the end of the arm, where the ring interlocked with the terminal ring, $\text{macro}_9$, exhibits a lower probability of engaging in intra-arm threading, disrupting the otherwise increasing trend. This reduced probability is likely due to both geometric and entropic constraints. Geometrically, $\text{macro}_9$ has fewer neighboring rings compared to those located deeper within the structure, limiting its interaction with the neighbors. Additionally, its position at the arm’s periphery restricts its freedom to adopt various configurations, imposing entropic constraints that further reduce the likelihood of threading. 

While the terminal ring, $\text{macro}_{10}$, experiences similar constraints, it shows a higher probability for serving as a host. This suggests that the terminal ring benefits from the increased freedom of movement due to its position at the edge, which allows it to interact more readily with its environment despite having fewer neighboring rings. In contrast to $\text{macro}_9$, the terminal ring's relative exposure and lack of crowding may make it more dynamic and accessible for threading events.

This effect may also be attributed to the number of available potential guests for each ring. To investigate this, the contribution of the arms to threading is first examined. Figure \ref{fig:inter_intra} (c) shows the distribution of threading events, decomposing the probability of events occurring within the same arm (referred to as 'same arm'), between two adjacent arms (referred to as 'first-nearest neighbors'), and between two most distant arms (referred to as 'second-nearest neighbors'). The data reveal that over 60\% of threading occurs within the same arm, between 30\% and 40\% occurs between adjacent arms, and less than 10\% occurs between distant arms, across all bending rigidities.

The index difference between the guest and host rings was analyzed for both inter-arm and intra-arm threading, defining $\textbf{index difference} = |I-J|$ in which $I$ is the index of host and $J$ is the index of guest in a threading event.
 The results indicate that over 90 \% of intra-arm threading occurs between two rings separated by only one ring, with $\textbf{index difference}=2$. This intermediate ring acts as a mediator, keeping the guest and the host rings in close proximity. Additionally, the mechanical bond provided by the intermediate ring facilitates their mobility, enhancing the likelihood of threading events.

For inter-arm threading, over 90\% of the events have an index difference of 0, meaning that the guest and host ring indices in two different arms are the same. Since threading between arms predominantly occurs through the $macro_1$, the ring interlocked with the central ring, this indicates that inter-arm threading primarily involves the interpenetration of the first rings of adjacent arms. This suggests that steric crowding plays a facilitating role in these interactions. The observation implies a mechanism similar to intra-arm threading, where the central ring mediates proximity and mobility, thus promoting threading interactions between the first rings of adjacent arms.

The star shape of the structure introduces two factors that enhance the probability of threading: steric crowding and conformational freedom along the arms. For inter-arm threading, steric crowding plays a facilitating role, particularly between the first rings of adjacent arms, where the proximity of rings near the central core promotes threading interactions. As the arms extend away from the center, steric hindrance diminishes, and the rings gain greater conformational freedom, which increases the likelihood of threading events.

At the ends of the arms, particularly for the rings labeled $\text{macro}_9$ and $\text{macro}_{10}$, these two factors—steric crowding and freedom— compete. While steric crowding near the center of the star enhances threading between neighboring arms, this effect weakens at the arm's periphery as the rings experience fewer constraints. The resulting balance between the decreased steric hindrance and increased flexibility may explain the sudden drop in threading probability at the end of the arm, particularly when $\text{macro}_9$ and $\text{macro}_{10}$ compete for involving in threading.

\subsection{Impact of steric crowding and arm flexibility on threading in star catenanes}

 \begin{figure}
    \centering
    \includegraphics[width=1.1\textwidth]{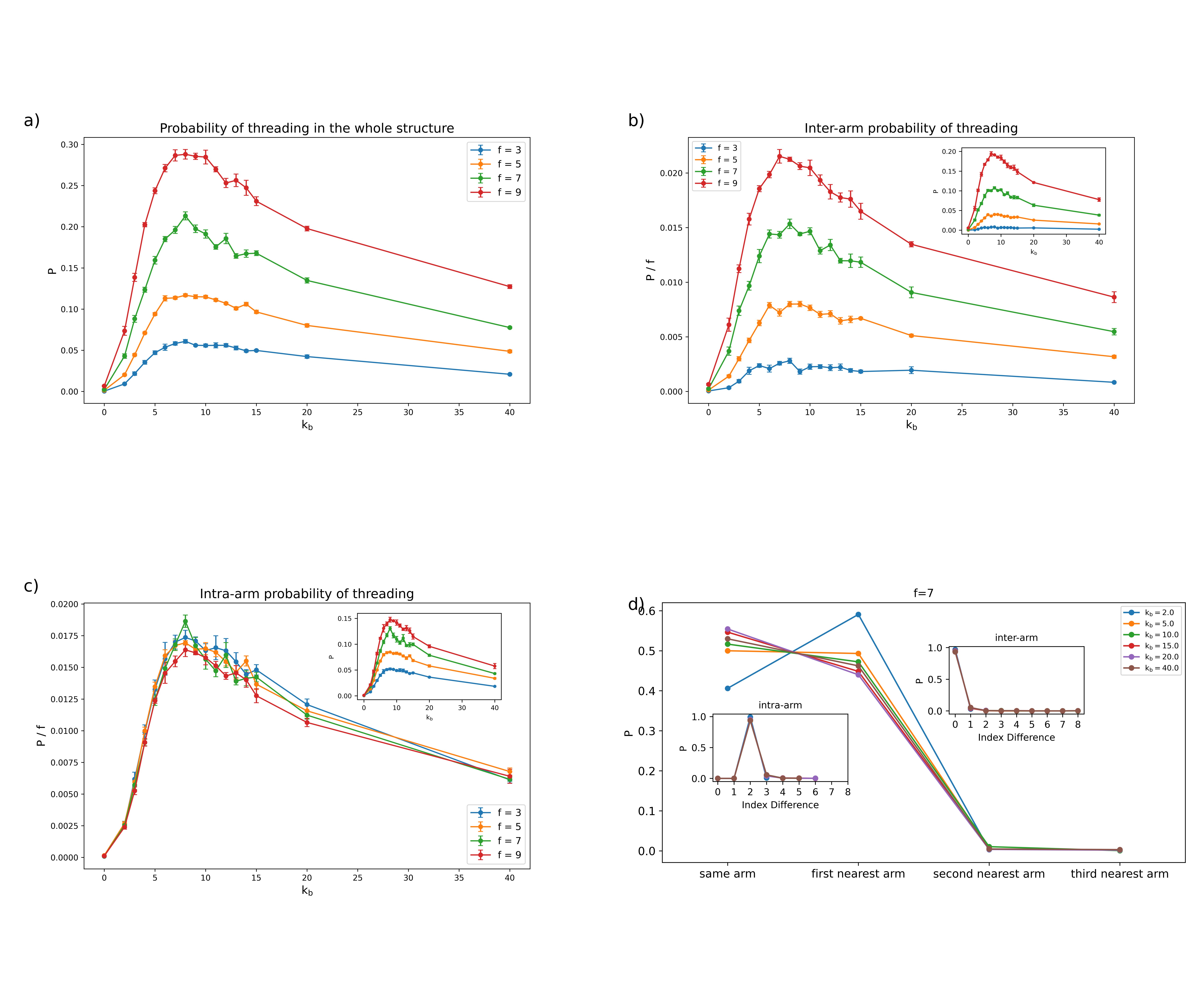}
    \caption{Analysis of Threading Probability in Star Catenanes with Varying Numbers of Arms. (a) Threading probability for star catenanes with different arm multiplicities (\(f=3, 5, 7, 9\)) as a function of bending rigidity. (b) Normalized inter-arm threading probability (\(P/f\)) for varying arm multiplicities. (c) Normalized intra-arm threading probability (\(P/f\)) for varying arm multiplicities. (d) Probability of inter- and intra-arm threading based on the relative positions of the arms accommodating the host and guest, categorized as same, first nearest, and second nearest arm. Subplots demonstrate the index differences between host and guest for intra- and inter-arm threading.}
    \label{fig:prob_f}
\end{figure}

To further investigate the impact of steric crowding and flexibility along the arms, alternative star catenane systems were examined with varying parameters compared to the reference system. Given that steric effects are associated with the degree of crowding at the star's center, star catenanes with fixed values of \(m=40\) and \(n=10\) were considered, while varying the arm multiplicity \(f\) as \(f=3, 5, 7, 9\). The overall threading probability is presented in Fig. \ref{fig:prob_f} (a). As observed, the threading probability exhibits a non-uniform trend, with a pronounced peak in the intermediate regime.

It is also noted that increasing \(f\) leads to an increase in the total number of rings in the system, which in turn raises the probability of threading. To examine how the increase in the number of arms—and the resulting crowding—affects the threading probability, inter-arm and intra-arm threading probabilities were separately analyzed. The contribution from central threading is negligible. 

As illustrated in Fig. \ref{fig:prob_f} (d), inter-arm threading predominantly occurs between the first two rings, while intra-arm threading involves two rings separated by a middle ring acting as a mediator. To neutralize the effect of the increasing number of rings due to increasing the number of arms, $f$, the threading probabilities are normalized by \(f\). 

The results show that intra-arm threading follows a similar trend and does not change significantly with increasing \(f\). In contrast, the crowding effect strongly influences inter-arm threading, leading to its increase \ref{fig:prob_f} (b). Notably, threading between two adjacent arms remains more probable even as \(f\) increases. This observation indicates that, in addition to the increasing number of rings associated with a higher \(f\), the central crowding effect significantly contributes to the total threading probability. This highlights the dual role of arm multiplicity in raising both the number of threading opportunities and steric crowding effect at the center.

 \begin{figure}
    \centering
    \includegraphics[width=1.1\textwidth]{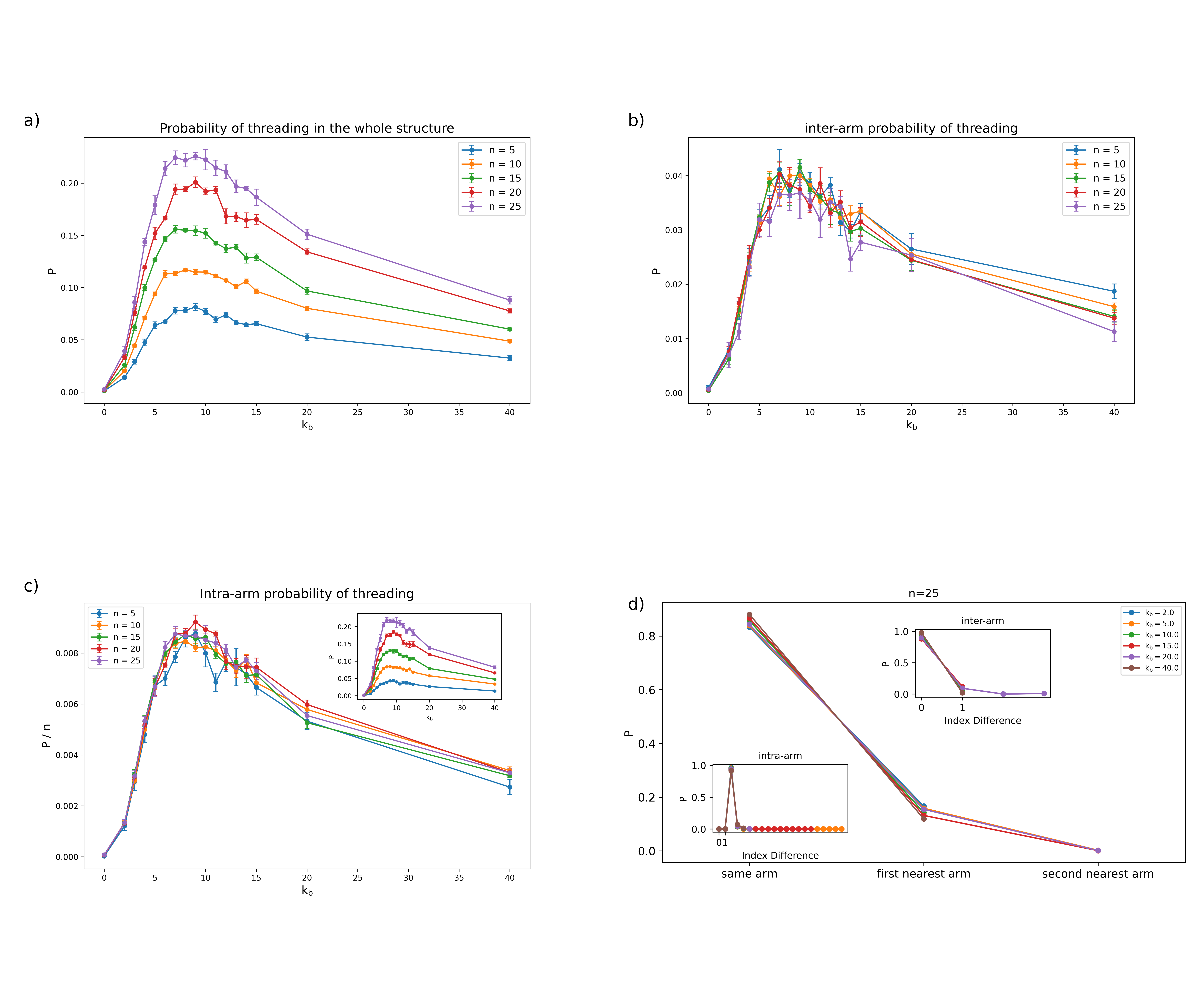}
    \caption{Analysis of Threading Probability in Star Catenanes with Varying Numbers of Arms. (a) Threading probability for star catenanes with different arm multiplicities (\(f=3, 5, 7, 9\)) as a function of bending rigidity. (b) Normalized inter-arm threading probability (\(P/f\)) for varying arm multiplicities. (c) Normalized intra-arm threading probability (\(P/f\)) for varying arm multiplicities. (d) Probability of inter- and intra-arm threading based on the relative positions of the arms accommodating the host and guest, categorized as same, first nearest, and second nearest arm. Subplots demonstrate the index differences between host and guest for intra- and inter-arm threading.}
    \label{fig:prob_n}
\end{figure}

Further, the star catenanes with \(m=40\), \(f=5\) and varying arm length \(n=5, 10, 15, 20, 25\) were analyzed to explore the impact of arm length on threading probability. As shown in Fig \ref{fig:prob_n} (a), as the arms become longer, the total threading probability increases, primarily due to the larger number of rings available for threading. However, normalizing the intra-arm threading probability by the number of rings in each arm (Fig. \ref{fig:prob_n} (c)) reveals a consistent trend across all systems, indicating that the increase in threading probability arises mainly from the increased number of rings rather than a change in threading dynamics within individual arms.

Given that the number of arms \(f=5\) is constant across the systems, the steric crowding at the center affects inter-arm threading similarly in all cases (Fig. \ref{fig:prob_n} (b)). One might expect that, with longer arms, inter-arm threading could involve rings beyond those directly interlocked with the central ring. However, this behavior was not observed; inter-arm threading remains restricted to rings directly connected to the central core as shown in \ref{fig:prob_n} (d). Additionally, the crowding effect at the core does not diminish with increased arm length.

These findings collectively demonstrate that while the flexibility and conformational freedom of longer arms might theoretically allow for more diverse threading interactions, threading events predominantly occur between next-nearest neighbors within a single arm or between rings closest to the central ring in inter-arm threading. Thus, the primary factor driving the increased probability of threading in these systems is the greater number of rings introduced by elongating the arms.

In conclusion, this study explores the probability of threading in star catenane systems, focusing on the interplay between the global topology of the star shape and the properties of the individual arms, which can be treated as linear catenanes. Using a reference system in which the number of arms and the arm length are fixed, the effects of bending rigidity and ring length on the threading probability were examined. The interaction between two length scales—the length of the rings and their persistence length due to rigidity—results in a non-uniform threading probability. This behavior arises because bending rigidity alters the ring shapes, leading to an uneven distribution of host and guest candidates for threading in intermediate stiffness. The results also indicate that increasing the length of the rings significantly boosts the probability of threading, which could be an effective strategy for enhancing threading in future applications.

Furthermore, the influence of star topology on threading dynamics was investigated. Within one arm, threading primarily occurs between next-nearest neighbors, with rings interlocked via a mediator ring. The presence of mechanical bonds, providing ring mobility, and the mediator, which maintains proximity between host and guest rings, facilitates threading within individual arms. Inter-arm threading predominantly involves the first rings of each arm, interlocked with the central ring, which acts as a mediator. These insights open avenues for future studies on the design of more complex catenane systems, potentially informing the development of advanced molecular machines and materials with tunable threading behaviors.
\begin{acknowledgement}

{\bf Acknowledgements} The financial support from  PNRR grant CN\_00000013\_CN-HPC, M4C2I1.4, spoke 7, funded by NextGenerationEU is acknowledged. Valuable suggestions regarding the identification of threading algorithm were provided by Cristian Micheletti, and his input is gratefully acknowledged.

\end{acknowledgement}


\bibliography{achemso-demo}

\providecommand{\latin}[1]{#1}
\makeatletter
\providecommand{\doi}
  {\begingroup\let\do\@makeother\dospecials
  \catcode`\{=1 \catcode`\}=2 \doi@aux}
\providecommand{\doi@aux}[1]{\endgroup\texttt{#1}}
\makeatother
\providecommand*\mcitethebibliography{\thebibliography}
\csname @ifundefined\endcsname{endmcitethebibliography}  {\let\endmcitethebibliography\endthebibliography}{}
\begin{mcitethebibliography}{37}
\providecommand*\natexlab[1]{#1}
\providecommand*\mciteSetBstSublistMode[1]{}
\providecommand*\mciteSetBstMaxWidthForm[2]{}
\providecommand*\mciteBstWouldAddEndPuncttrue
  {\def\EndOfBibitem{\unskip.}}
\providecommand*\mciteBstWouldAddEndPunctfalse
  {\let\EndOfBibitem\relax}
\providecommand*\mciteSetBstMidEndSepPunct[3]{}
\providecommand*\mciteSetBstSublistLabelBeginEnd[3]{}
\providecommand*\EndOfBibitem{}
\mciteSetBstSublistMode{f}
\mciteSetBstMaxWidthForm{subitem}{(\alph{mcitesubitemcount})}
\mciteSetBstSublistLabelBeginEnd
  {\mcitemaxwidthsubitemform\space}
  {\relax}
  {\relax}

\bibitem[Hoagland(1997)]{Hoagland-1997}
Hoagland,~D. The physics of polymers: Concepts for understanding their structures and behavior, by Gert R. Strobl, Springer-Verlag, New York, 1996. ISBN 3-540-60768-4. \emph{Journal of Polymer Science Part A: Polymer Chemistry} \textbf{1997}, \emph{35}, 1337--1338\relax
\mciteBstWouldAddEndPuncttrue
\mciteSetBstMidEndSepPunct{\mcitedefaultmidpunct}
{\mcitedefaultendpunct}{\mcitedefaultseppunct}\relax
\EndOfBibitem
\bibitem[Rubinstein and Colby(2003)Rubinstein, and Colby]{rubinstein2003polymer}
Rubinstein,~M.; Colby,~R. \emph{Polymer Physics}; Oxford University Press, 2003\relax
\mciteBstWouldAddEndPuncttrue
\mciteSetBstMidEndSepPunct{\mcitedefaultmidpunct}
{\mcitedefaultendpunct}{\mcitedefaultseppunct}\relax
\EndOfBibitem
\bibitem[McLeish \latin{et~al.}(1999)McLeish, Allgaier, Bick, Bishko, Biswas, Blackwell, Blottière, Clarke, Gibbs, Groves, Hakiki, Heenan, Johnson, Kant, Read, and Young]{McLeish-1999}
McLeish,~T. C.~B. \latin{et~al.}  Dynamics of Entangled H Polymers: Theory, Rheology, and Neutron-Scattering. \emph{Macromolecules} \textbf{1999}, \emph{32}, 6734--6758\relax
\mciteBstWouldAddEndPuncttrue
\mciteSetBstMidEndSepPunct{\mcitedefaultmidpunct}
{\mcitedefaultendpunct}{\mcitedefaultseppunct}\relax
\EndOfBibitem
\bibitem[Gil-Ramírez \latin{et~al.}(2015)Gil-Ramírez, Leigh, and Stephens]{GilRamirez-2015}
Gil-Ramírez,~G.; Leigh,~D.~A.; Stephens,~A.~J. Catenanes: Fifty Years of Molecular Links. \emph{Angewandte Chemie International Edition} \textbf{2015}, \emph{54}, 6110--6150\relax
\mciteBstWouldAddEndPuncttrue
\mciteSetBstMidEndSepPunct{\mcitedefaultmidpunct}
{\mcitedefaultendpunct}{\mcitedefaultseppunct}\relax
\EndOfBibitem
\bibitem[Tubiana \latin{et~al.}(2024)Tubiana, Alexander, Barbensi, Buck, Cartwright, Chwastyk, Cieplak, Coluzza, Čopar, Craik, {Di Stefano}, Everaers, Faísca, Ferrari, Giacometti, Goundaroulis, Haglund, Hou, Ilieva, Jackson, Japaridze, Kaplan, Klotz, Li, Likos, Locatelli, López-León, Machon, Micheletti, Michieletto, Niemi, Niemyska, Niewieczerzal, Nitti, Orlandini, Pasquali, Perlinska, Podgornik, Potestio, Pugno, Ravnik, Ricca, Rohwer, Rosa, Smrek, Souslov, Stasiak, Steer, Sułkowska, Sułkowski, Sumners, Svaneborg, Szymczak, Tarenzi, Travasso, Virnau, Vlassopoulos, Ziherl, and Žumer]{TUBIANA-2024}
Tubiana,~L. \latin{et~al.}  Topology in soft and biological matter. \emph{Physics Reports} \textbf{2024}, \emph{1075}, 1--137, Topology in soft and biological matter\relax
\mciteBstWouldAddEndPuncttrue
\mciteSetBstMidEndSepPunct{\mcitedefaultmidpunct}
{\mcitedefaultendpunct}{\mcitedefaultseppunct}\relax
\EndOfBibitem
\bibitem[Sikorski(1994)]{SIKORSKI_1994}
Sikorski,~A. Monte Carlo study of catenated ring polymers. \emph{Polymer} \textbf{1994}, \emph{35}, 3792--3794\relax
\mciteBstWouldAddEndPuncttrue
\mciteSetBstMidEndSepPunct{\mcitedefaultmidpunct}
{\mcitedefaultendpunct}{\mcitedefaultseppunct}\relax
\EndOfBibitem
\bibitem[Grosberg \latin{et~al.}(1988)Grosberg, Nechaev, and Shakhnovich]{grosberg-1988}
Grosberg,~A.~Y.; Nechaev,~S.; Shakhnovich,~E. {The role of topological constraints in the kinetics of collapse of macromolecules}. \emph{{Journal de Physique}} \textbf{1988}, \emph{49}, 2095--2100\relax
\mciteBstWouldAddEndPuncttrue
\mciteSetBstMidEndSepPunct{\mcitedefaultmidpunct}
{\mcitedefaultendpunct}{\mcitedefaultseppunct}\relax
\EndOfBibitem
\bibitem[Wu \latin{et~al.}(2017)Wu, Rauscher, Lang, Wojtecki, de~Pablo, Hore, and Rowan]{Wu_2017}
Wu,~Q.; Rauscher,~P.~M.; Lang,~X.; Wojtecki,~R.~J.; de~Pablo,~J.~J.; Hore,~M. J.~A.; Rowan,~S.~J. Poly[<i>n</i>]catenanes: Synthesis of molecular interlocked chains. \emph{Science} \textbf{2017}, \emph{358}, 1434--1439\relax
\mciteBstWouldAddEndPuncttrue
\mciteSetBstMidEndSepPunct{\mcitedefaultmidpunct}
{\mcitedefaultendpunct}{\mcitedefaultseppunct}\relax
\EndOfBibitem
\bibitem[Halverson \latin{et~al.}(2012)Halverson, Grest, Grosberg, and Kremer]{Halverson-2012}
Halverson,~J.~D.; Grest,~G.~S.; Grosberg,~A.~Y.; Kremer,~K. Rheology of Ring Polymer Melts: From Linear Contaminants to Ring-Linear Blends. \emph{Phys. Rev. Lett.} \textbf{2012}, \emph{108}, 038301\relax
\mciteBstWouldAddEndPuncttrue
\mciteSetBstMidEndSepPunct{\mcitedefaultmidpunct}
{\mcitedefaultendpunct}{\mcitedefaultseppunct}\relax
\EndOfBibitem
\bibitem[Vargas-Lara \latin{et~al.}(2018)Vargas-Lara, Pazmiño~Betancourt, and Douglas]{Vargas-2018}
Vargas-Lara,~F.; Pazmiño~Betancourt,~B.~A.; Douglas,~J.~F. Communication: A comparison between the solution properties of knotted ring and star polymers. \emph{The Journal of Chemical Physics} \textbf{2018}, \emph{149}, 161101\relax
\mciteBstWouldAddEndPuncttrue
\mciteSetBstMidEndSepPunct{\mcitedefaultmidpunct}
{\mcitedefaultendpunct}{\mcitedefaultseppunct}\relax
\EndOfBibitem
\bibitem[Ahmadian~Dehaghani \latin{et~al.}(2020)Ahmadian~Dehaghani, Chubak, Likos, and Ejtehadi]{Ahmadian-2020}
Ahmadian~Dehaghani,~Z.; Chubak,~I.; Likos,~C.~N.; Ejtehadi,~M.~R. Effects of topological constraints on linked ring polymers in solvents of varying quality. \emph{Soft Matter} \textbf{2020}, \emph{16}, 3029--3038\relax
\mciteBstWouldAddEndPuncttrue
\mciteSetBstMidEndSepPunct{\mcitedefaultmidpunct}
{\mcitedefaultendpunct}{\mcitedefaultseppunct}\relax
\EndOfBibitem
\bibitem[Farimani \latin{et~al.}(2024)Farimani, Ahmadian~Dehaghani, Likos, and Ejtehadi]{Farimani-2024}
Farimani,~R.~A.; Ahmadian~Dehaghani,~Z.; Likos,~C.~N.; Ejtehadi,~M.~R. Effects of Linking Topology on the Shear Response of Connected Ring Polymers: Catenanes and Bonded Rings Flow Differently. \emph{Phys. Rev. Lett.} \textbf{2024}, \emph{132}, 148101\relax
\mciteBstWouldAddEndPuncttrue
\mciteSetBstMidEndSepPunct{\mcitedefaultmidpunct}
{\mcitedefaultendpunct}{\mcitedefaultseppunct}\relax
\EndOfBibitem
\bibitem[Chen \latin{et~al.}(2024)Chen, Xu, Rao, and Zhang]{Chen-2024}
Chen,~Y.; Xu,~D.; Rao,~Y.; Zhang,~G. Nonlinear Elasticity of Single Linear Polycatenane: Emergence of Stress-Softening. \emph{Macromolecules} \textbf{2024}, \emph{57}, 9041--9058\relax
\mciteBstWouldAddEndPuncttrue
\mciteSetBstMidEndSepPunct{\mcitedefaultmidpunct}
{\mcitedefaultendpunct}{\mcitedefaultseppunct}\relax
\EndOfBibitem
\bibitem[Tubiana \latin{et~al.}(2022)Tubiana, Ferrari, and Orlandini]{Luca_2022}
Tubiana,~L.; Ferrari,~F.; Orlandini,~E. Circular Polycatenanes: Supramolecular Structures with Topologically Tunable Properties. \emph{Phys. Rev. Lett.} \textbf{2022}, \emph{129}, 227801\relax
\mciteBstWouldAddEndPuncttrue
\mciteSetBstMidEndSepPunct{\mcitedefaultmidpunct}
{\mcitedefaultendpunct}{\mcitedefaultseppunct}\relax
\EndOfBibitem
\bibitem[Dai and Doyle(2016)Dai, and Doyle]{Dai-2016}
Dai,~L.; Doyle,~P.~S. Effects of Intrachain Interactions on the Knot Size of a Polymer. \emph{Macromolecules} \textbf{2016}, \emph{49}, 7581--7587\relax
\mciteBstWouldAddEndPuncttrue
\mciteSetBstMidEndSepPunct{\mcitedefaultmidpunct}
{\mcitedefaultendpunct}{\mcitedefaultseppunct}\relax
\EndOfBibitem
\bibitem[Rusková and Račko(2023)Rusková, and Račko]{Ruskova-2023}
Rusková,~R.; Račko,~D. Knot Formation on DNA Pushed Inside Chiral Nanochannels. \emph{Polymers} \textbf{2023}, \emph{15}\relax
\mciteBstWouldAddEndPuncttrue
\mciteSetBstMidEndSepPunct{\mcitedefaultmidpunct}
{\mcitedefaultendpunct}{\mcitedefaultseppunct}\relax
\EndOfBibitem
\bibitem[Tubiana \latin{et~al.}(2013)Tubiana, Rosa, Fragiacomo, and Micheletti]{Tubiana-2013}
Tubiana,~L.; Rosa,~A.; Fragiacomo,~F.; Micheletti,~C. Spontaneous Knotting and Unknotting of Flexible Linear Polymers: Equilibrium and Kinetic Aspects. \emph{Macromolecules} \textbf{2013}, \emph{46}, 3669--3678\relax
\mciteBstWouldAddEndPuncttrue
\mciteSetBstMidEndSepPunct{\mcitedefaultmidpunct}
{\mcitedefaultendpunct}{\mcitedefaultseppunct}\relax
\EndOfBibitem
\bibitem[Suma \latin{et~al.}(2015)Suma, Rosa, and Micheletti]{suma-2015}
Suma,~A.; Rosa,~A.; Micheletti,~C. Pore Translocation of Knotted Polymer Chains: How Friction Depends on Knot Complexity. \emph{ACS Macro Letters} \textbf{2015}, \emph{4}, 1420--1424, PMID: 35614794\relax
\mciteBstWouldAddEndPuncttrue
\mciteSetBstMidEndSepPunct{\mcitedefaultmidpunct}
{\mcitedefaultendpunct}{\mcitedefaultseppunct}\relax
\EndOfBibitem
\bibitem[Coronel \latin{et~al.}(2017)Coronel, Orlandini, and Micheletti]{Coronel-2017-Softmatter}
Coronel,~L.; Orlandini,~E.; Micheletti,~C. Non-monotonic knotting probability and knot length of semiflexible rings: the competing roles of entropy and bending energy. \emph{Soft Matter} \textbf{2017}, \emph{13}, 4260--4267\relax
\mciteBstWouldAddEndPuncttrue
\mciteSetBstMidEndSepPunct{\mcitedefaultmidpunct}
{\mcitedefaultendpunct}{\mcitedefaultseppunct}\relax
\EndOfBibitem
\bibitem[Wang \latin{et~al.}(2024)Wang, Lu, and Sun]{WANG-2024}
Wang,~W.; Lu,~J.; Sun,~R. Effect of threading on static and dynamic properties of polymer chains in entangled linear ring blend systems with different stiffness. \emph{Polymer} \textbf{2024}, \emph{290}, 126513\relax
\mciteBstWouldAddEndPuncttrue
\mciteSetBstMidEndSepPunct{\mcitedefaultmidpunct}
{\mcitedefaultendpunct}{\mcitedefaultseppunct}\relax
\EndOfBibitem
\bibitem[Rosa \latin{et~al.}(2020)Rosa, Smrek, Turner, and Michieletto]{rosa-2020}
Rosa,~A.; Smrek,~J.; Turner,~M.~S.; Michieletto,~D. Threading-Induced Dynamical Transition in Tadpole-Shaped Polymers. \emph{ACS Macro Letters} \textbf{2020}, \emph{9}, 743--748, PMID: 33828901\relax
\mciteBstWouldAddEndPuncttrue
\mciteSetBstMidEndSepPunct{\mcitedefaultmidpunct}
{\mcitedefaultendpunct}{\mcitedefaultseppunct}\relax
\EndOfBibitem
\bibitem[Tu \latin{et~al.}(2023)Tu, Davydovich, Mei, Singh, Grest, Schweizer, O’Connor, and Schroeder]{Tu-2023}
Tu,~M.~Q.; Davydovich,~O.; Mei,~B.; Singh,~P.~K.; Grest,~G.~S.; Schweizer,~K.~S.; O’Connor,~T.~C.; Schroeder,~C.~M. Unexpected Slow Relaxation Dynamics in Pure Ring Polymers Arise from Intermolecular Interactions. \emph{ACS Polymers Au} \textbf{2023}, \emph{3}, 307--317\relax
\mciteBstWouldAddEndPuncttrue
\mciteSetBstMidEndSepPunct{\mcitedefaultmidpunct}
{\mcitedefaultendpunct}{\mcitedefaultseppunct}\relax
\EndOfBibitem
\bibitem[Michieletto \latin{et~al.}(2014)Michieletto, Marenduzzo, Orlandini, Alexander, and Turner]{Micheletto-2014-Macroletter}
Michieletto,~D.; Marenduzzo,~D.; Orlandini,~E.; Alexander,~G.~P.; Turner,~M.~S. Threading Dynamics of Ring Polymers in a Gel. \emph{ACS Macro Letters} \textbf{2014}, \emph{3}, 255--259, PMID: 35590516\relax
\mciteBstWouldAddEndPuncttrue
\mciteSetBstMidEndSepPunct{\mcitedefaultmidpunct}
{\mcitedefaultendpunct}{\mcitedefaultseppunct}\relax
\EndOfBibitem
\bibitem[Michieletto and Turner(2016)Michieletto, and Turner]{Micheletto-2016-PNAs}
Michieletto,~D.; Turner,~M.~S. A topologically driven glass in ring polymers. \emph{Proceedings of the National Academy of Sciences} \textbf{2016}, \emph{113}, 5195--5200\relax
\mciteBstWouldAddEndPuncttrue
\mciteSetBstMidEndSepPunct{\mcitedefaultmidpunct}
{\mcitedefaultendpunct}{\mcitedefaultseppunct}\relax
\EndOfBibitem
\bibitem[Lee \latin{et~al.}(2015)Lee, Kim, and Jung]{Lee-Rapid}
Lee,~E.; Kim,~S.; Jung,~Y. Slowing Down of Ring Polymer Diffusion Caused by Inter-Ring Threading. \emph{Macromolecular Rapid Communications} \textbf{2015}, \emph{36}, 1115--1121\relax
\mciteBstWouldAddEndPuncttrue
\mciteSetBstMidEndSepPunct{\mcitedefaultmidpunct}
{\mcitedefaultendpunct}{\mcitedefaultseppunct}\relax
\EndOfBibitem
\bibitem[Smrek \latin{et~al.}(2019)Smrek, Kremer, and Rosa]{Smrek-2019-Macroletter}
Smrek,~J.; Kremer,~K.; Rosa,~A. Threading of Unconcatenated Ring Polymers at High Concentrations: Double-Folded vs Time-Equilibrated Structures. \emph{ACS Macro Letters} \textbf{2019}, \emph{8}, 155--160, PMID: 30800531\relax
\mciteBstWouldAddEndPuncttrue
\mciteSetBstMidEndSepPunct{\mcitedefaultmidpunct}
{\mcitedefaultendpunct}{\mcitedefaultseppunct}\relax
\EndOfBibitem
\bibitem[Chubak \latin{et~al.}(2020)Chubak, Likos, Kremer, and Smrek]{Chubak-2020}
Chubak,~I.; Likos,~C.~N.; Kremer,~K.; Smrek,~J. Emergence of active topological glass through directed chain dynamics and nonequilibrium phase segregation. \emph{Phys. Rev. Res.} \textbf{2020}, \emph{2}, 043249\relax
\mciteBstWouldAddEndPuncttrue
\mciteSetBstMidEndSepPunct{\mcitedefaultmidpunct}
{\mcitedefaultendpunct}{\mcitedefaultseppunct}\relax
\EndOfBibitem
\bibitem[Ubertini \latin{et~al.}(2022)Ubertini, Smrek, and Rosa]{Ubertini-2022}
Ubertini,~M.~A.; Smrek,~J.; Rosa,~A. Entanglement Length Scale Separates Threading from Branching of Unknotted and Non-concatenated Ring Polymers in Melts. \emph{Macromolecules} \textbf{2022}, \emph{55}, 10723--10736\relax
\mciteBstWouldAddEndPuncttrue
\mciteSetBstMidEndSepPunct{\mcitedefaultmidpunct}
{\mcitedefaultendpunct}{\mcitedefaultseppunct}\relax
\EndOfBibitem
\bibitem[Guo \latin{et~al.}(2020)Guo, Li, Wu, He, and Zhang]{Guo-2020-polymers}
Guo,~F.; Li,~K.; Wu,~J.; He,~L.; Zhang,~L. Effects of Topological Constraints on Penetration Structures of Semi-Flexible Ring Polymers. \emph{Polymers} \textbf{2020}, \emph{12}\relax
\mciteBstWouldAddEndPuncttrue
\mciteSetBstMidEndSepPunct{\mcitedefaultmidpunct}
{\mcitedefaultendpunct}{\mcitedefaultseppunct}\relax
\EndOfBibitem
\bibitem[Dehaghani \latin{et~al.}(2023)Dehaghani, Chiarantoni, and Micheletti]{Ahmadian-2023}
Dehaghani,~Z.~A.; Chiarantoni,~P.; Micheletti,~C. Topological Entanglement of Linear Catenanes: Knots and Threadings. \emph{ACS Macro Letters} \textbf{2023}, \emph{12}, 1231--1236, PMID: 37638542\relax
\mciteBstWouldAddEndPuncttrue
\mciteSetBstMidEndSepPunct{\mcitedefaultmidpunct}
{\mcitedefaultendpunct}{\mcitedefaultseppunct}\relax
\EndOfBibitem
\bibitem[Staňo \latin{et~al.}(2023)Staňo, Likos, and Smrek]{Stano-2023-Softmatter}
Staňo,~R.; Likos,~C.~N.; Smrek,~J. To thread or not to thread? Effective potentials and threading interactions between asymmetric ring polymers. \emph{Soft Matter} \textbf{2023}, \emph{19}, 17--30\relax
\mciteBstWouldAddEndPuncttrue
\mciteSetBstMidEndSepPunct{\mcitedefaultmidpunct}
{\mcitedefaultendpunct}{\mcitedefaultseppunct}\relax
\EndOfBibitem
\bibitem[Thompson \latin{et~al.}(2022)Thompson, Aktulga, Berger, Bolintineanu, Brown, Crozier, in~'t Veld, Kohlmeyer, Moore, Nguyen, Shan, Stevens, Tranchida, Trott, and Plimpton]{LAMMPS}
Thompson,~A.~P.; Aktulga,~H.~M.; Berger,~R.; Bolintineanu,~D.~S.; Brown,~W.~M.; Crozier,~P.~S.; in~'t Veld,~P.~J.; Kohlmeyer,~A.; Moore,~S.~G.; Nguyen,~T.~D.; Shan,~R.; Stevens,~M.~J.; Tranchida,~J.; Trott,~C.; Plimpton,~S.~J. {LAMMPS} - a flexible simulation tool for particle-based materials modeling at the atomic, meso, and continuum scales. \emph{Comp. Phys. Comm.} \textbf{2022}, \emph{271}, 108171\relax
\mciteBstWouldAddEndPuncttrue
\mciteSetBstMidEndSepPunct{\mcitedefaultmidpunct}
{\mcitedefaultendpunct}{\mcitedefaultseppunct}\relax
\EndOfBibitem
\bibitem[Kremer and Grest(1990)Kremer, and Grest]{kremer_1990}
Kremer,~K.; Grest,~G.~S. Dynamics of entangled linear polymer melts: A molecular dynamics simulation. \emph{The Journal of Chemical Physics} \textbf{1990}, \emph{92}, 5057--5086\relax
\mciteBstWouldAddEndPuncttrue
\mciteSetBstMidEndSepPunct{\mcitedefaultmidpunct}
{\mcitedefaultendpunct}{\mcitedefaultseppunct}\relax
\EndOfBibitem
\bibitem[Deledalle \latin{et~al.}(2017)Deledalle, Denis, Tabti, and Tupin]{eigen}
Deledalle,~C.-A.; Denis,~L.; Tabti,~S.; Tupin,~F. \emph{{Closed-form expressions of the eigen decomposition of 2 x 2 and 3 x 3 Hermitian matrices}}; Research Report, 2017\relax
\mciteBstWouldAddEndPuncttrue
\mciteSetBstMidEndSepPunct{\mcitedefaultmidpunct}
{\mcitedefaultendpunct}{\mcitedefaultseppunct}\relax
\EndOfBibitem
\bibitem[Alim and Frey(2007)Alim, and Frey]{Alim-2007}
Alim,~K.; Frey,~E. Shapes of Semiflexible Polymer Rings. \emph{Phys. Rev. Lett.} \textbf{2007}, \emph{99}, 198102\relax
\mciteBstWouldAddEndPuncttrue
\mciteSetBstMidEndSepPunct{\mcitedefaultmidpunct}
{\mcitedefaultendpunct}{\mcitedefaultseppunct}\relax
\EndOfBibitem
\bibitem[Rauscher \latin{et~al.}(2020)Rauscher, Schweizer, Rowan, and de~Pablo]{Rauscher-2020}
Rauscher,~P.~M.; Schweizer,~K.~S.; Rowan,~S.~J.; de~Pablo,~J.~J. Thermodynamics and Structure of Poly[n]catenane Melts. \emph{Macromolecules} \textbf{2020}, \emph{53}, 3390--3408\relax
\mciteBstWouldAddEndPuncttrue
\mciteSetBstMidEndSepPunct{\mcitedefaultmidpunct}
{\mcitedefaultendpunct}{\mcitedefaultseppunct}\relax
\EndOfBibitem
\end{mcitethebibliography}
\end{document}